 \documentclass[twocolumn,trackchanges]{aastex62} %linenumbers
\graphicspath{{./}{figures/}}

\received{Jul. 31st, 2018}
\revised{Aug. 2nd, 2018}
\accepted{Sep. 18th, 2018}
\submitjournal{ApJL}

\shorttitle{Time Stamps of Vertical Phase Mixing in the Disk}
\shortauthors{Tian et al.}
\def\kms{\,\mathrm{km\,s}^{-1}}
\newcommand{\teff}{${T}_{\rm eff}$}
\newcommand {\Usun}{{U_{\!\odot}}}
\newcommand {\Wsun}{{W_{\!\odot}}}
\newcommand {\Vsun}{{V_{\!\odot}}}

\newcommand{\vR}{$\langle v_R\rangle$}

\newcommand{\vPhi}{$\langle v_\phi\rangle$}

\begin{document}

\title{Time stamps of vertical phase mixing in the Galactic disk from LAMOST-Gaia stars}

\correspondingauthor{Haijun Tian, Chao Liu}
\email{hjtian@lamost.org, liuchao@nao.cas.cn}

\author[0000-0001-9289-0589]{Hai-Jun Tian}
\affil{China Three Gorges University, Yichang 443002, China}
\affil{Center of Astronomy and Space Science Research, China Three Gorges University, Yichang 443002, China}
\affil{Max Planck Institute for Astronomy, K\"onigstuhl 17, D-69117 Heidelberg, Germany}

\author[0000-0002-1802-6917]{Chao Liu}
\author{Yue Wu}
\author{Mao-Sheng Xiang}
\affiliation{Key Lab for Optical Astronomy, National Astronomical Observatories, Chinese Academy of Sciences, Beijing 100012, China}
\author{Yong ZHANG}
\affiliation{Nanjing Institute of Astronomical Optics \& Technology, National Astronomical Observatories, Chinese Academy of Sciences, Nanjing 210042, China}
%\collaboration{(LaTeX collaboration)}

%% Note that the \and command from previous versions of AASTeX is now
%% depreciated in this version as it is no longer necessary. AASTeX 
%% automatically takes care of all commas and "and"s between authors names.

%% AASTeX 6.2 has the new \collaboration and \nocollaboration commands to
%% provide the collaboration status of a group of authors. These commands 
%% can be used either before or after the list of corresponding authors. The
%% argument for \collaboration is the collaboration identifier. Authors are
%% encouraged to surround collaboration identifiers with ()s. The 
%% \nocollaboration command takes no argument and exists to indicate that
%% the nearby authors are not part of surrounding collaborations.

%% Mark off the abstract in the ``abstract'' environment. 
\begin{abstract}
The perturbation mechanism of the Galactic disk has puzzled us for a long time. The imprints from perturbations provide important diagnostics on the disk formation and evolution. Here we try to constrain when the vertical perturbation took place in the disk by tracking the phase mixing history. Firstly, we clearly depict the spiral structures of radial ($v_R$) and azimuthal ($v_{\phi}$) velocities in the phase space of the vertical position and velocity ($z$-$v_z$) with 723,871 LAMOST-Gaia combined stars. Then, we investigate the variation of the spirals with stellar age ($\tau$) by dividing the sample into seven stellar age bins. Finally, we find that the spirals explicitly exist in all the bins, even in the bin of $\tau<0.5$\,Gyr, except for the bin of $\tau>6.0$\,Gyr. This constrains the vertical perturbation probably starting no later than 0.5\,Gyr ago. But we can not rule out whether the young stars ($\tau<0.5$\,Gyr) inherit the oscillations from the perturbed ISM where they born from. This study provides some important observational evidences to understand the disk perturbation mechanisms, even the formation and evolution of our Galaxy.

%Whereas the results in this study are important to check the basic assumption whether the young stars inherit the oscillations from the perturbed ISM where the young stars born from, and for the perturbation mechanism constraint, but more observational informations or hydrodynamical simulations in high resolution are required.

%Besides, some other features about the phase spiral structures are summarized in this Letter. These findings provide vital observational evidences to understand the disk perturbation mechanisms, even the formation and evolution of our Galaxy.

%If the phase mixing is sparked by an external accretion event, e.g. the passage of the Sagittarius dwarf galaxy, we can infer that the Sagittarius dwarf galaxy passed by the Galactic disk in the last 0.4\,Gyr which is tightly constrained relative to the time between 0.3 and 0.9\,Gyr claimed by \citet{antoja2018}. But the results can not rule out the other perturbation induced by the Galactic bar and/or spiral structures. 

\end{abstract}

%% Keywords should appear after the \end{abstract} command. 
%% See the online documentation for the full list of available subject
%% keywords and the rules for their use.

%% From the front matter, we move on to the body of the paper.
%% Sections are demarcated by \section and \subsection, respectively.
%% Observe the use of the LaTeX \label
%% command after the \subsection to give a symbolic KEY to the
%% subsection for cross-referencing in a \ref command.
%% You can use LaTeX's \ref and \label commands to keep track of
%% cross-references to sections, equations, tables, and figures.
%% That way, if you change the order of any elements, LaTeX will
%% automatically renumber them.
%%
%% We recommend that authors also use the natbib \citep
%% and \citet commands to identify citations.  The citations are
%% tied to the reference list via symbolic KEYs. The KEY corresponds
%% to the KEY in the \bibitem in the reference list below. 

\section{Introduction} \label{sec:intro}
%April 25th of 2018 is a memorable date for the astronomy community. At that day, Gaia delivered its second data release (DR2). More than 1.3 billion stars are measured the precise positions, proper motions and parallax, of which 7.2 million bright stars are also measured the line-of-sight velocity. Fruitful outputs have been made from the data (Katz et al. 2018; Antoja et al. 2018; Kawata et al. 2018). 

The Galactic disk has been revealed to suffer from perturbations, which have imprints in both the stellar density \citep{widrow2012, gmo+13, xu2015, wang2018b, xiang2018} and kinematics \citep{siebert2011, tian2015, tian2017, wang2018a, gaia2018a}. The perturbations in the plane are typically explained in terms of the influence of the non-axisymmetric components of the Galaxy, e.g. the rotating bar \citep{monari2013, bovy2015}, the spiral arms \citep{siebert2012, faure2014, kawata2014} or both \citep{quillen2011, grand2015, tian2017}. The vertical perturbations are usually thought to be excited by minor mergers, such as the passage of the Sagittarius dwarf galaxy through the Galactic disc \citep{gmo+13}. However, some proofs were found from simulations that the rotating bar and the spiral arms can also give rise to significant vertical influence \citep{monari2013, faure2014, debattista2014}. \citet{tian2015} showed that the asymmetric motion may be related to the age of the stars, i.e. the younger populations show larger peculiar velocity in both radial and vertical directions than the old populations. Also the non-zero radial flow can be simply explained with the projection effect in the elliptical stellar orbits \citep{tian2017}.

More recently, \citet[hereafter A18]{antoja2018} found, from Gaia DR2 stars, that the disk is full of substructures in the phase space. It indicates that the Galactic disk is currently undergoing the phase mixing, and unveils that the Galactic disk has experienced vertical perturbation. In the phase space of $z$-$v_z$, the stars take on an impressive curled spiral-shaped distribution in the radial velocity ($v_R$) and the azimuthal velocity ($v_{\phi}$). Based on the impulse-approximation, \citet{binney2018} built a simple model to explain the origin of the phase-plane spiral. The key factors that lead to a spiral are: that the vertical frequency $\Omega_z$ depends on angular momentum not only in $J_z$ but also in $J_\phi$, and that the stellar vertical oscillations should be in anharmonic state, which could be perturbed by an intruder, such as a dwarf galaxy or a pure dark-matter structure. \citet{laporte2018} used a set of numerical N-body simulations, in which a Sgittarius-like dSph (Sgr) hits a Milky Way-like host, to follow the orbit of Sgr from the first pericentric passage to the present-day, and illustrate the evolution of the phase-space spirals in the last Gyr.
%The authors well reproduced the spiral shape in a simulation, which supports the perturbation of minor mergers. Finally, the authors claimed that the Sagittarius dwarf galaxy might be the culprit to bring the disk to a non-equilibrium state, due to its strong effects on the stellar disk between 0.3 and 0.9\,Gyr ago \citep{purcell2011, gomez2012}.

In this study, we propose to constrain when the perturbation took place in the disk, by tracking the phase mixing history with different stellar population. The cornerstone Gaia has already measured precise proper motions and distances for more than 1.3 billion stars. And the LAMOST survey \citep{zhao2012,liu2014} has collected more than 9 million stellar spectra, and built the largest stellar atmospheric parameter library in the world. The combination of the Gaia and LAMOST provides us an unprecedented sample to track the history of the phase mixing in a large spatial volume.

This Letter is organized as follows. In Section \ref{sec:data_selection}, the sample selection is briefly described. The results and discussion are presented in Section \ref{sec:results}. Finally, we conclude this work in Section \ref{sec:conclusion}.

Throughout the paper, we adopt the solar motion as $(\Usun,\Vsun,\Wsun) = (9.58, 10.52, 7.01)\kms$\, \citep{tian2015}, the circular speed of the local standard of rest (LSR) as $v_0=238\kms$ \citep{schonrich2012}, and the solar Galactocentric radius and vertical positions as ($R_0$, $z_0$) = (8.27,0.02)\,kpc \citep{schonrich2010,schonrich2012}. 

%According to the strength of the spiral pattern in different stellar population, one can track the history of the phase mixing. A18 can not investigate the variation of the phase substructure with the stellar populations since Gaia DR2 is lack of the stellar atmospherical parameters. LAMOST Survey, operated by the National Astronomical Observatories, Chinese Academy of Sciences, has delivered millions of spectra, which have been derived the stellar astrophysical parameters (effective temperature, surface gravity, and metallicity) as well as line-of-sight velocities. With these parameters, it is easy to obtain the different stellar populations. Combining the LAMOST and Gaia data, we can construct a currently best sample for the research on the phase mixing. We intend to figure out when, where and how the perturbation happened to the phase mixing with a serial of papers.

%Siebert et al. (2012) applied two-dimensional density wave to analytically model the radial component of the two-dimensional velocity field, and well explain the radial velocity gradient detected in Siebert et al. 2011; 
%Antoja et al. (2018) present remarkable observational evidences that  

\section{The Sample Selection} \label{sec:data_selection}
In order to build the sample containing stellar astrophysical parameters and precise kinematical informations, we combine the data from the two surveys: the LAMOST spectroscopic survey \citep{zhao2012, cui2012}, and Gaia \citep{gaia2016} survey. 

LAMOST is a quasi-meridian reflecting Schmidt telescope with an effective aperture of about 4\,m and a field of view of 5\degr. The LAMOST Survey has internally delivered the fifth data release (DR5), which contains 9,017,844 spectra with a resolution of $\sim$1800 covering a wavelength range of 3800\,\AA$ \lesssim \lambda \lesssim$9000\,\AA. In the catalog, the LAMOST stellar parameter pipeline - LASP has derived those stellar astrophysical parameters (\teff, $\log g$, and [Fe/H]) as well as the line-of-sight velocities for 5,475,513 stars \citep{wu2011a,wu2014}. LAMOST radial velocities are as precise as about 5 $\kms$, but with a systematical under-estimation of $\sim$ 5.7 $\kms$\citep{tian2015}.% in the $\sim$1000 cross-identified LAMOST-APOGEE stars.  %schonrich2017

Gaia DR2 \citep{gaia2018} includes more than 1.3 billion stars in the Milk Way with G-band magnitude brighter than $\sim$20.7, which were measured the precise position, proper motions and parallaxes. \citet{Bailer-Jones2018} calculated the distances and the asymmetric uncertainties for the stars from the parallaxes in the Gaia DR2. %Gaia is a cornerstone mission conducted by the European Space Agency (ESA). 

\begin{figure}[!t]
\centering
\includegraphics[width=0.46\textwidth, trim=0.0cm 0.0cm 1.cm 0.0cm, clip]{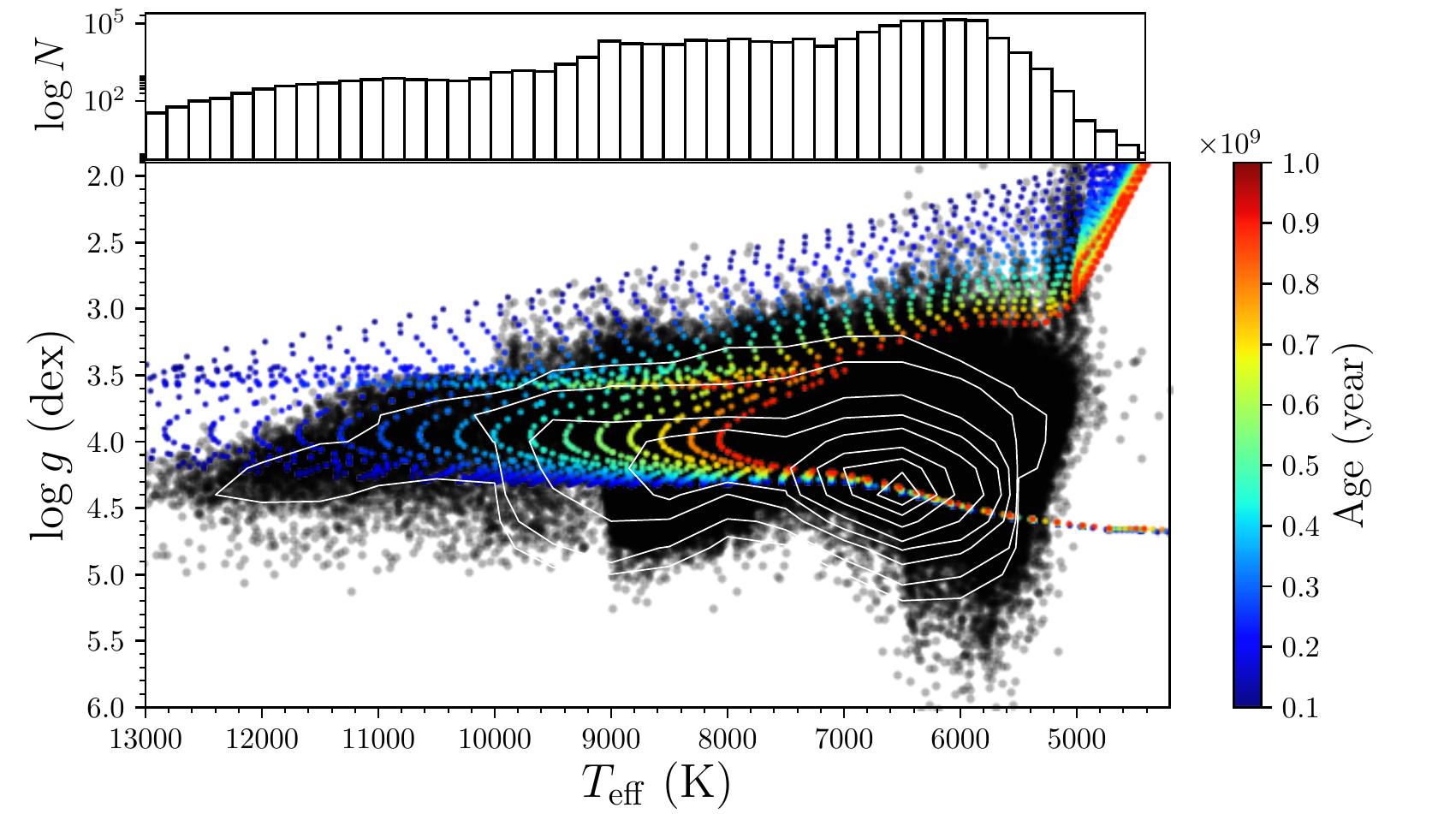}
\caption{The sample distribution in the \teff-$\log g$ panel. The black points display the 723,871 LAMOST-Gaia stars used in this work. The isochrones are color-coded with stellar ages, to demonstrate the distribution of young stellar population ($\tau<1.0$\,Gyr). The histogram in the top sub-panel presents the logarithmic number of stars in \teff\ bins. The contours indicate the normalized number density of stars with different levels of 0.003, 0.01, 0.05, 0.1, 0.2, 0.4, 0.6, and 0.8 (the highest density is normalized to 1). % and the green unfilled histogram is same but for the MSTO sample from Xiang et al. (2015), which are used for the validation in Section \ref{sec:validation}.
}\label{fig:teff_logg}
\end{figure}

\begin{figure*}[!t]
\centering
\includegraphics[width=0.3\textwidth, trim=0.0cm 0.0cm 0.0cm 0.0cm, clip]{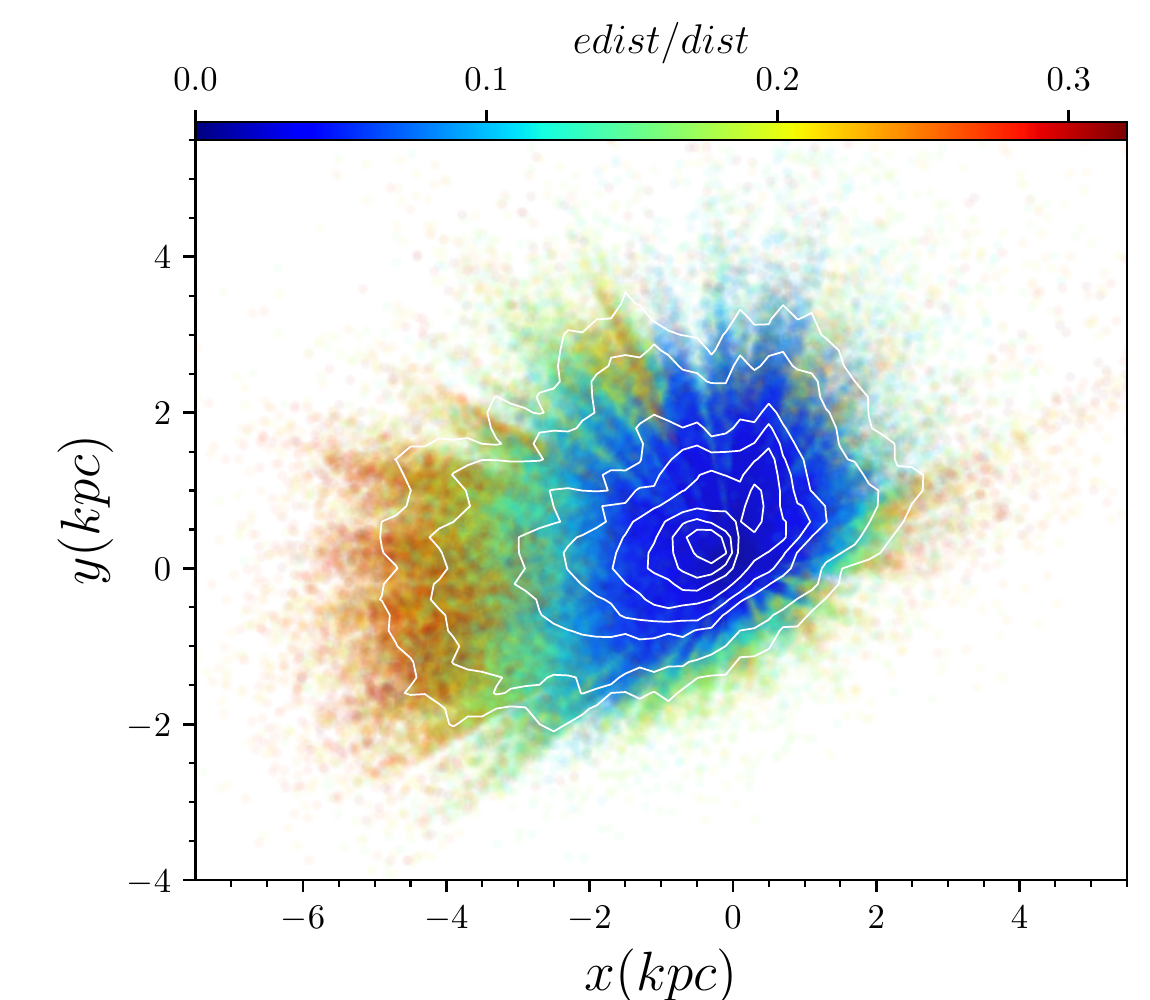}
\includegraphics[width=0.3\textwidth, trim=0.0cm 0.0cm 0.0cm 0.0cm, clip]{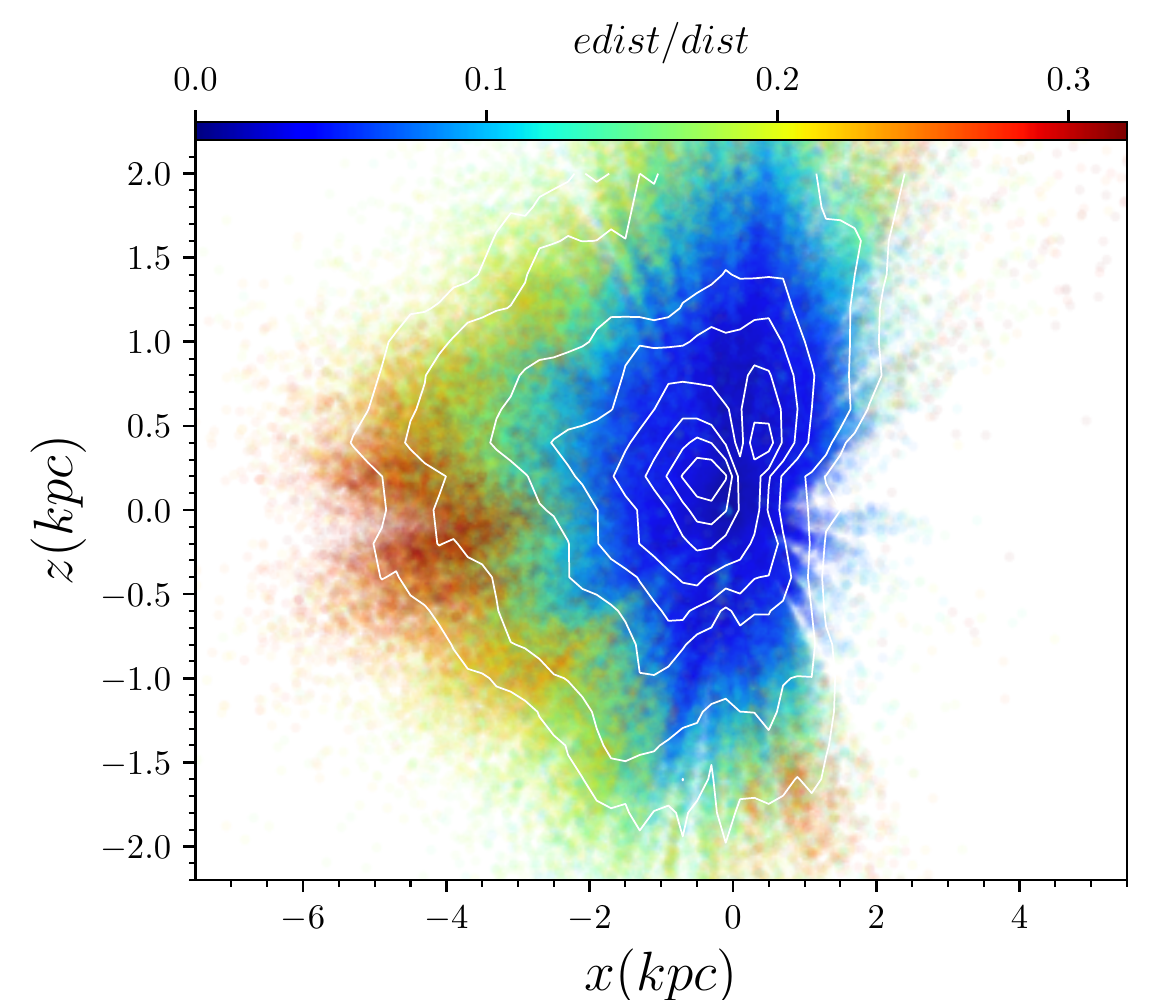}
\includegraphics[width=0.3\textwidth, trim=0.0cm 0.0cm 0.0cm 0.0cm, clip]{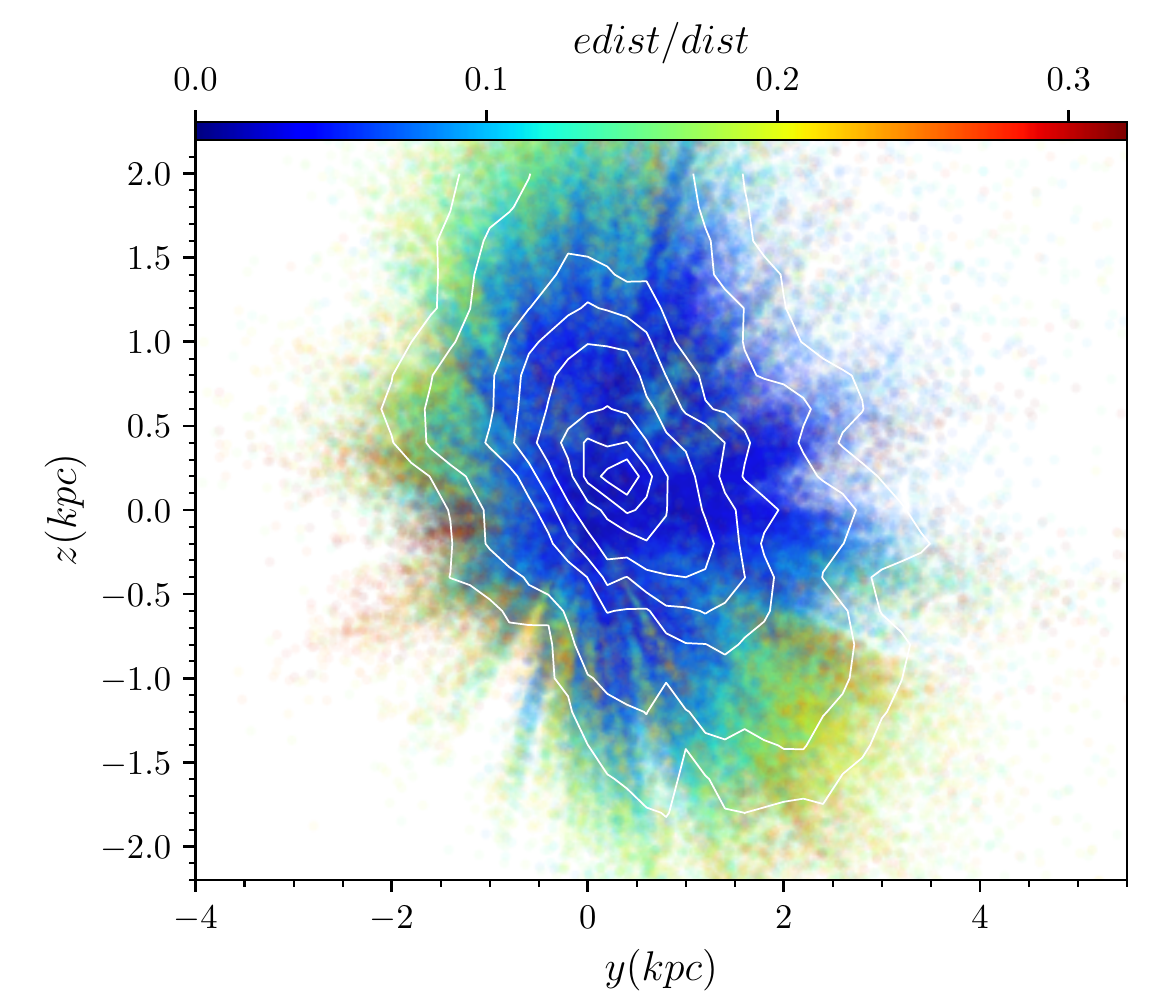}
\caption{The spatial distribution of the sample in the Cartesian coordinates. The left (x-y), middle (x-z), and right (y-z) panels show the projected stellar distributions, respectively. The color bars indicate the relative distance errors measured from Gaia DR2 parallaxes by \citet{Bailer-Jones2018}. In total, more than 95\% stars have relative distance errors ($edist/dist$) of $<20\%$. The contours have the same meaning as Fig. \ref{fig:teff_logg}.
}\label{fig:xyz}
\end{figure*}

Cross-matching the LAMOST DR5 with Gaia DR2, and removing the duplicated observations, we obtained 4,130,116 stars which not only have the LAMOST astrophysical parameters and line-of-sight velocity, but also have Gaia distances and proper motions. This provides us an unprecedented sample to track the stellar phase mixing history.

In oder to obtain the stellar ages, we further cross-matched the LAMOST-Gaia sample with an updated-version of the age catalog from \citet{xiang2017}. The catalog of \citet{xiang2017} contains stellar ages and masses of 0.93 million Galactic-disk main-sequence turnoff (MSTO) and sub-giant stars from LAMOST DR4. The ages are determined by matching with stellar isochrones using a Bayesian algorithm, utilizing effective temperature \teff, absolute magnitude $M_V$, metallicity [Fe/H], and $\alpha$-element to iron abundance ratio [$\alpha$/Fe] deduced from the LAMOST spectra. In this work, the stellar ages are re-calculated with the same procedure of \citet{xiang2017}, but using the absolute magnitudes deduced from Gaia DR2 parallaxes. To avoid strong bias in the deduced absolute magnitudes (thus ages) from the Gaia parallaxes, we restrict our sample stars with relative parallax errors smaller than 20\%. The precision of the newly estimated ages are significantly improved relative to that of \citet{xiang2017}. The whole sample has a median error of 22\% for the age estimates, and about 70\% of the stars are older than 2\,Gyr. %have a median age uncertainties of 21\%. %Removing the stars without Gaia distances, we obtain 895, 950 stars.  % but for the young stars (<1.0 Gyr), the age uncertainty is around 40\%

We notice that \citet{xiang2017} discard all stars with \teff$>10000$\, K, because the atmospheric parameters of those hot stars given by LSP3 \citep{xiang2015} have poor precision. The temperature cut discards most stars younger than 0.5\,Gyr, which are extremely important for the purpose of current paper.

To select younger stars, we pre-select A-type star candidates according to the following empirical criteria \citep{liu2015}, 
\begin{itemize}\itemsep1pt
\item $EW_{\rm Fe} < 0.6\ \&\ EW_{\rm H\gamma}<6\ \&\ EW_{\rm Fe}<(0.6-0.2)/(6-13)*(EW_{\rm H\gamma}-6)+0.6\ \&\ EW_{\rm H\gamma}>=6$,
\item $EW_{\rm G4300}<0.2\ \&\ EW_{\rm H\gamma}<6\ \&\ G<(0.2+2.5)/(6-13.2)*(EW_{\rm H\gamma}-6)+0.2\ \&\ EW_{\rm H\gamma}>=6\ \&\ EW_{\rm Mg}<(-0.2-1.2)/(-5-16.5)*(EW_{\rm H\gamma}+5)-0.2$,
\end{itemize}
where $EW_{\rm H\gamma}$ and $EW_{\rm G4300}$ are equivalent widths (EWs) of $H_{\gamma}$ and G-band (defined in \citet{liu2015}). $EW_{\rm Fe}$ is the averaged EW of nine Fe lines located at 4383, 4531, 4668, 5015, 5270, 5335, 5406, 5709 and 5782 \AA, $EW_{\rm Mg}$ is the averaged EW of three Lick indices, i.e. Mg I, Mg II and Mg${_b} $\citep{worthey1994}. 

Then we derive the atmospheric parameters for the A-type star candidates by adopting ULySS package \citep{wu2011b}, and estimate the age for each star with the method in \citet{xiang2017}. The A-type stars used in this work are further purified with the following criteria:
\begin{itemize}\itemsep1pt
    \item \teff$>$7500 K,
    \item $\log{g} >$ 3.5,
    \item SNR$ > 30$ in g-band
\end{itemize}

To get a sample with good age estimation, we discard stars with relative age error of $>33\%$. Finally, we obtain a sample of 723,871 LAMOST-Gaia stars, shown as the black dots in Fig. \ref{fig:teff_logg}. The color-coded isochrones with stellar ages mainly display the distribution of young stars ($\tau<1.0$\,Gyr) in the \teff-$\log{g}$ panel. The spatial distributions of the sample in the Cartesian coordinate system are displayed in Fig \ref{fig:xyz}, each dot is color-coded with its relative distance error. In total, more than 95\% stars have relative distance error of $<20\%$. %in which 680,561 MSTO and subgiant stars from \citet{xiang2017}, and 43,310 newly selected stars are contained,  %in total 927,453

\section{Results and Discussion} \label{sec:results}
We divide the sample into seven stellar age ($\tau$) bins from 0.2\,Gyr to $\sim$10.0\,Gyr, to investigate the spiral patterns at different stellar ages. Each stellar bin includes at least hundreds of thousands stars. The stellar number ($N$) of each bin is labeled in Fig. \ref{fig:z_vz_teffbins}. To reduce the contamination of old stars, the youngest stellar bin with $\tau<$0.5\,Gyr only includes A-type stars. % i.e. \teff$>7500$\, K.

%\begin{table*}
%\caption{ The stellar numbers in the seven stellar age bins}.\label{tab:bin_num}
%\centering
%\begin{tabular}{c|c|c|c|c|c|c|c}
%\hline
%\hline
%Age(Gyr)&$\tau<0.4$&$0.4<\tau<0.6$&$0.6<\tau<0.8$&$0.8<\tau<1.0$&$1.0<\tau<2.0$&$2.0<\tau<6.0$&$\tau>6.0$\\
%\hline
%Num&18,837&42,294&33,697&35,337&128,949&414,115&218,357\\
%\hline
%\hline
%\end{tabular}\\
%\end{table*}

\subsection{The spirals in the LAMOST-Gaia Sample}\label{sec:spiral}
\begin{figure}[t]
\centering
\includegraphics[width=0.25\textwidth, trim=0.0cm 1.3cm 0.03cm 1.4cm, clip]{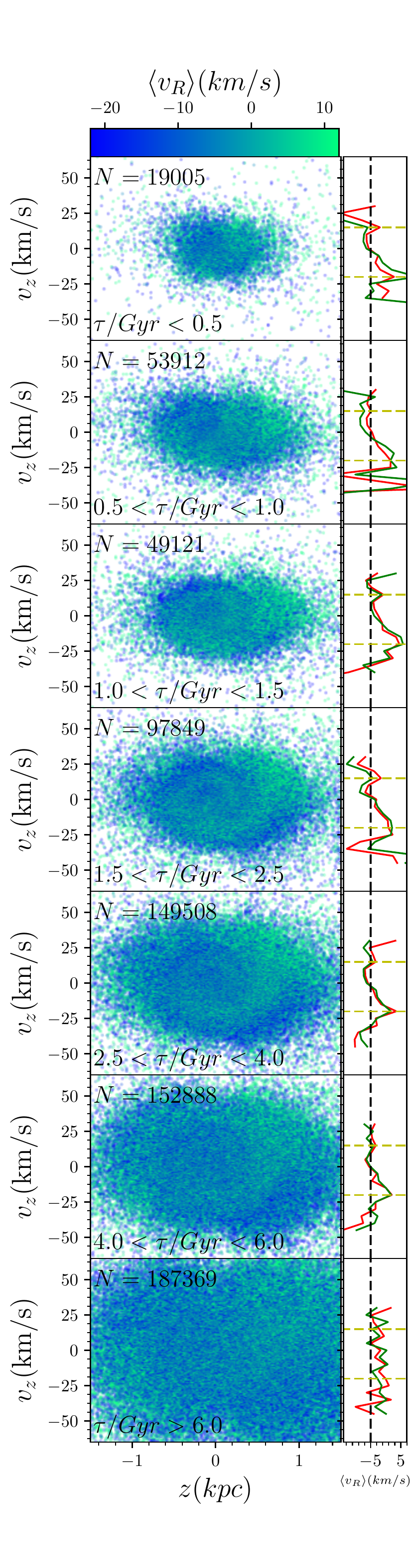} %0.25
\includegraphics[width=0.1962\textwidth, trim=1.77cm 1.3cm 0.05cm 1.4cm, clip]{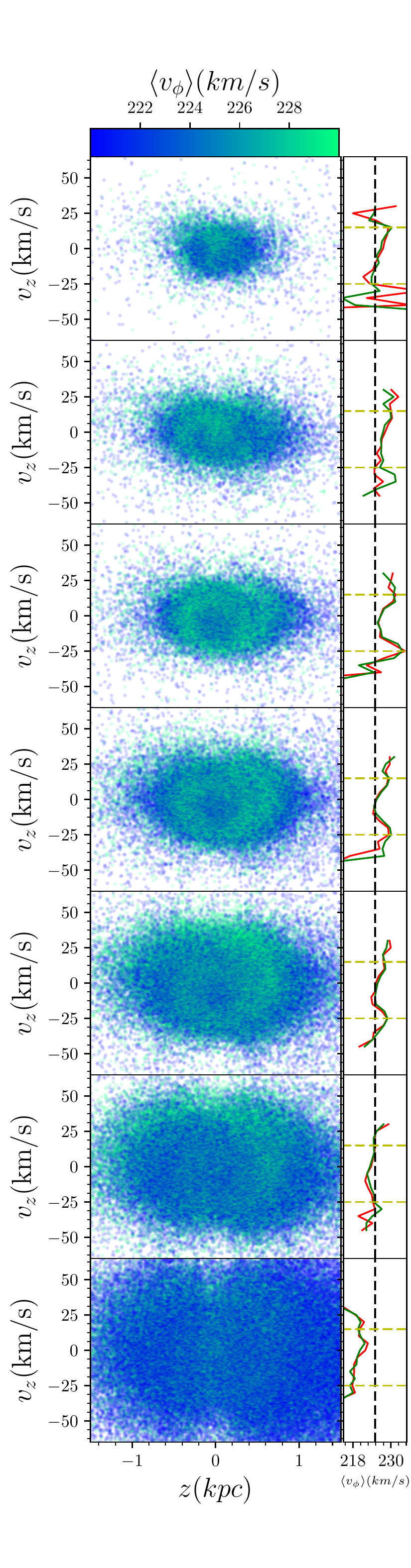} %0.1962
\caption{The spiral-shaped structures in the phase space of the vertical position and velocity ($z$-$v_z$) for the LAMOST-Gaia stars. The left panel is colored as a function of median $v_R$. The right panel is colored as a function of median $v_\phi$. The pixel has a size of $\Delta z$ = 0.01\,kpc and $\Delta v_z$ = 0.1 $\kms$. Here $N$ means the star number in each age bin. The red and green curves in the minor sub-panels demonstrate the variations of the \vR\ (or \vPhi) with $v_z$ in the two slices of $|z|<0.1$\,kpc and $|z+0.1|<0.1$\,kpc. The yellow dashed lines roughly indicate the locations of peaks in the \vR\ and \vPhi\ curves. The black dashed lines are just for reference, which are \vR\ = -5 $\kms$ and \vPhi\ = 225 $\kms$, respectively.
}\label{fig:z_vz_teffbins}
\end{figure}

We reproduce the spiral-shaped structures in the phase space of $z$-$v_z$ from the LAMOST-Gaia stars. Fig. \ref{fig:z_vz_teffbins} displays the $z$-$v_z$ phase-space spirals for \vR\ (the left column) and \vPhi\ (the right column) in different stellar age ($\tau$) bins. The pixel has a size of $\Delta z$ = 0.01\,kpc and $\Delta v_z$ = 0.1 $\kms$. The red and green curves in the minor sub-panels demonstrate the variations of \vR\ (or \vPhi) with $v_z$ in the two slices of $|z|<0.1$\,kpc and $|z+0.1|<0.1$\,kpc, respectively. The yellow dashed lines mark the locations of $v_z$ = 15.0 $\kms$ and $v_z$ = -20.0 $\kms$ in the \vR\ sub-panel, and the locations of $v_z$ = 15.0$\kms$ and $v_z$ = -25.0$\kms$ in the \vPhi\ sub-panel. They are roughly marked the locations of peaks in the \vR\ and \vPhi\ curves. 

The spirals are significantly prominent for  both \vR\ and \vPhi, in particular in the middle three age bins, i.e. $1.0\,\rm{Gyr}<\tau<1.5$\,Gyr (the 3rd row), $1.5\,\rm{Gyr}<\tau<2.5$\,Gyr (the 4th row), and $2.5\,\rm{Gyr}<\tau<4.0$\,Gyr (the 5th row). In the first and last second bins, i.e. $0.5\,\rm{Gyr}<\tau<1.0$\,Gyr and $4.0\,\rm{Gyr}<\tau<6.0$\,Gyr, the spirals for both \vR\ and \vPhi\ are not clear, but still visible. The spirals in $\tau<0.5$\,Gyr are indistinguishable, since young stellar population is kinematically cold, the stars are confined to the centre of the $z$-$v_z$ plane by the age-velocity dispersion relation. As the stellar age growing, however, the spirals become obscure in $\tau>4.0$\,Gyr. When the stellar ages become larger than 6\,Gyr, the spirals are almost disappeared. It indicates that the old stars, which are the kinematically hot stellar population, are not sensitive to the perturbation.

The red and green curves in the minor sub-panels show clear wiggles for both \vR\ and \vPhi\ in all the age bins, even in $\tau<0.5$\,Gyr, but except for $\tau>6.0$\,Gyr. And each wiggle has two pronounced peaks, which indicate two wraps in the $z$-$v_z$ phase space. The interesting thing is that the peaks for \vR\  (or \vPhi) are corresponding to the same $v_z$ values in each age bin, and the peaks of \vR\ and \vPhi\ are roughly located in the same $v_z$ values, except the 5\,$\kms$ difference at the negative $v_z$ peak. These similar wiggles suggest that the spirals probably have the same origin. %In the slices at $|v_z|\sim 0 \kms$, \vR\ and \vPhi\ have similar but slightly weaker wiggles than those in the slices with $|z| \sim 0$\,kpc. 

%The astrophysical parameters including \teff$,\log{g}$, line-of-sight velocity, etc. of all the stars in the sample, are independently calculated with another pipeline LSP3 (Xiang et al. 2015), so it is befitting for the validation.

%\begin{figure}[t]
%\centering
%\includegraphics[width=0.25\textwidth, trim=0.0cm 0.0cm 0.05cm 0.0cm, clip]{z_VZ_VR_LMGaia_Agebins.png}
%\includegraphics[width=0.1962\textwidth, trim=1.78cm 0.0cm 0.0cm 0.0cm, clip]{z_VZ_VPhi_LMGaia_Agebins.png}
%\caption{Same as Fig. \ref{fig:z_vz_teffbins}, but for the sample from \citet{xiang2017} divided in the stellar age bins. Because the \teff\ of this sample is truncated at 10000 K, this sample is difficult to trace the stars younger than 1.0 Gyr, according to the isochrones displayed in Fig. \ref{fig:teff_logg}. But it is a good sample to trace when the spiral structure is disappeared in the old stellar population. 
%}\label{fig:z_vz_agebins}
%\end{figure}\includegraphics[]{ms.pdf}

\subsection{A possible starting time of the phase mixing}\label{sec:stime}
We suppose that the on-going vertical phase mixing is sparked by an external perturbation, which not only impacts the stars, but also pulls the ISM coherently as it does the stars. If the perturbation on the ISM can be quickly dissipated, the newly born stars from the perturbed gas will not share the general oscillations. According to the age of the stellar population, in which the $z$-$v_z$ phase spirals just appear, we could deduce the starting time of the phase mixing.

The most prominent feature in Fig. \ref{fig:z_vz_teffbins} is that the spirals are gradually apparent from $\tau<0.5$\,Gyr, and then slowly disappearing until $\tau>6.0$\,Gyr. This gradual changes of the spirals indicate that the different stellar populations have diverse sensitivity level to the perturbation. In addition, the oscillation signals have already existed in the stars with $\tau<0.5$\,Gyr, as shown by the 1-D red and green curves in the minor sub-panels. This suggests that the vertical perturbation to the disk probably took place no later than 0.5\,Gyr ago. This is consistent with the time predicted by A18, in which the authors claimed that the vertical phase mixing event started about 500\,Myr ago, with an uncertainty range of [300, 900]\,Myr. This is a strong constraint on the perturbation time, but it is under the assumption that the young stars do not carry on the perturbation contribution from the stimulated ISM where they are born from. It may be an ideal hypothesis. We will discuss more about it in Section \ref{sec:dis}.

\subsection{Discussion}\label{sec:dis}
%Besides the spiral pattern, Antoja et al. (2018) also observed many other substructures in the phase space distribution of stars in the disk of the Milky Way, such as arches and shells in velocity space, and snail shells and diagonal ridges in spatial and velocity combined phase space. These substructures indicate that the the disk is phase mixing from an out of equilibrium state.

%Disk stars usually are modeled as classic harmonic oscillators to describe their vertical movement in the epicyclic theory \citep{binney2008}. If the disk stars have same vertical oscillatory frequency, the phase mixing will not happen in the assumption of small amplitude. But when the disk stars are disturbed by a perturbation, the frequencies of stars will be slightly different. As time goes by, a spiral structure will be stretched out in the phase space \citep{tremaine1999, candlish2013}. It is called vertical phase mixing or wrapping model. 

%The stellar age is one of the most important quantities from the observational data in this work. 
The explanations to the spirals in Section \ref{sec:stime} are concise, but controversial. In this Section, we will discuss the issues in three aspects.
\subsubsection{Clues of an early perturbation history}\
There are several clues, which favor that the vertical perturbation to the disk took place in $\tau<0.5$\,Gyr ago:
\begin{itemize}\itemsep1pt
    \item Except for the old population in $\tau>6.0$\,Gyr, the spirals markedly existed in all the stellar age bins, especially in $\tau>0.5$\,Gyr. It suggests that stars in the disk probably have already been perturbed at least one time in the last 0.5\,Gyr. This is a natural and straightforward explanation.
     \item The pictures of \vR\ and \vPhi\ in  $\tau<0.5$\,Gyr look similar with the scenes presented by \citet{laporte2018} in Fig. 5 and Fig. 6 at $t\sim0$\,Gyr, during which a Sagittarius-like dSph passed by the pericentre.  
     \item The spirals in the different stellar bins look similar, and only take on two wraps, even in the old population in $4.0<\tau<6.0$\,Gyr.  The footprints of the dSph in the N-body simulation illustrate that the spirals can be formed in $\sim0.4$\,Gry after the Sgr's last pericenric passage (see Fig. 5 and Fig. 6 of \citet{laporte2018}). This also can be found in Fig. 7 of \citet{binney2018}. If the perturbation took place  $\tau>0.5$\,Gyr ago, the spirals perhaps will have more than two wraps, as showed in Fig. 9 of \citet{darling2018}.
\end{itemize}

\subsubsection{Doubts on the perturbation time}\
Although some clues support an early perturbation history, but there still exist some debatable points: 
\begin{itemize}\itemsep1pt
    \item Is the perturbation on ISM able to dissipated in short time? This is an open question. If not, it suggests that for all the stars, whatever they were born before or after the perturbation took place, both $v_R$ and $v_\phi$ will take on spiral patterns in the $z$-$v_z$ phase space, or at least include the perturbation contribution in each stellar age bin. Strictly speaking, we can not rule out this case in this work.
  \item Stars possibly oscillate much more slowly than one usually image in the vertical direction. It suggests that the spirals possibly can not be formed within $\sim0.4$\,Gry after the perturbation in reality.   %by using the frequencies at which stars would oscillate in a fixed $\phi$. But actually the coherent oscillations of the disc that a intruder excites, will case $\phi$ to change materially. 
    \item The perturbation time in different literatures are not converged well. \citet{minchev2009} predicted that the Galactic disk was strongly perturbed $\sim$1.9\,Gyr ago, and associated the perturbation with Galactic bar formation. \citet{monari2018} claimed that the disk experienced a vertical perturbation by the Sagittarius dwarf galaxy $\sim$1.5\, Gyr ago, and related it to the formation of the Coma Berenices moving group. A18 also inferred that the disk perturbation is caused by the passage of Sagittarius dwarf galaxy, but it occurred between 300 and 900\,Myr ago. In this study, some clues suggest that the vertical perturbation in the disk possibly took place no later than 0.5\,Gyr ago.  %The sub-structures discovered in the Gaia DR2 data, e.g. arches and shells in $v_R$ and $v_{\phi}$ velocity distribution, and ridges in spatial-velocity diagram \citep{gaia2018a,antoja2018,kawata2018,quillen2018}, are well explained with spiral arms \citep{kawata2018,quillen2018}, the rotating bar \citep{antoja2018}, and horizontal phase mixing model \citep{minchev2009, antoja2018}. Howerver,
\end{itemize}

\subsubsection{The comparison of stellar ages with others}
\begin{figure*}[!t]
\centering
\includegraphics[width=0.999\textwidth, trim=0.0cm 0.0cm 0.0cm 0.0cm, clip]{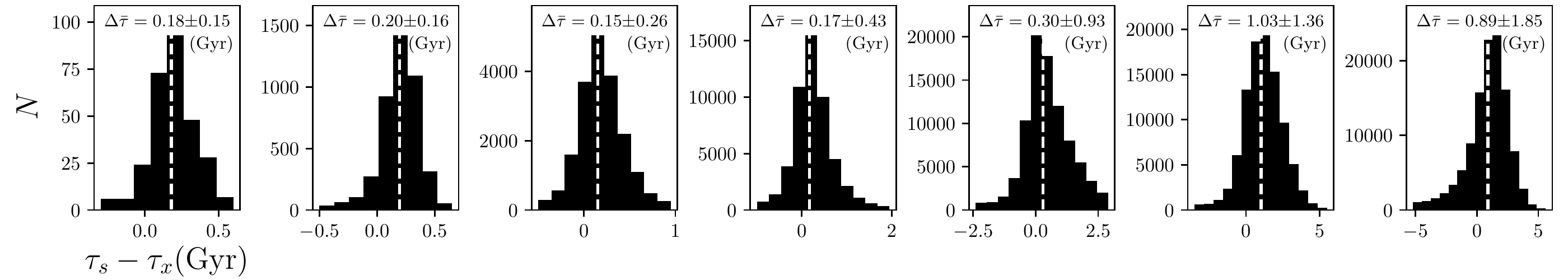}
\caption{ The histograms of age differences ($\Delta \tau = \tau_s - \tau_x$) of 416,140 cross-matched stars between our ages ($\tau_x$) and the ages ($\tau_s$) from \citet{Sanders2017}. The stars are divided into seven ages bins (ages increasing from the left to right) adopted in Fig. \ref{fig:z_vz_teffbins}. The white dashed lines mark the locations of the median age difference ($\Delta\bar\tau$). The stellar ages are consistent within 1$\sigma$, except in the first two young stellar bins. However, $\tau_s$ is systematically larger than $\tau_x$ by 8\% $\sim$ 30\% in general. % and the green unfilled histogram is same but for the MSTO sample from Xiang et al. (2015), which are used for the validation in Section \ref{sec:validation}.
}\label{fig:age_comp}
\end{figure*}

\
The results in Section \ref{sec:stime} might be affected by the bias of stellar ages. In order to investigate how the stellar ages are determined in this work, we compare our ages ($\tau_x$) with the ages ($\tau_s$) from \citet{Sanders2017} based on the 416,140 cross-matched stars. Fig. \ref{fig:age_comp} display the histograms of age differences ($\Delta \tau = \tau_s - \tau_x$) in the seven stellar age bins adopted in Fig. \ref{fig:z_vz_teffbins}. 

As one can see, the two stellar ages are consistent within 1$\sigma$, except in the first two young stellar bins. However, $\tau_s$ is systematically larger than $\tau_x$ by 10\% in $\tau>1.0$\,Gyr, and by $\sim$ 30\% in $\tau<1.0$\,Gyr. The systematical differences between $\tau_x$ and $\tau_s$ are  probably due to the effect of [$\alpha$/Fe]. We used [$\alpha$/Fe] to determine the age for each star, but \citet{Sanders2017} did not. And we found that the parameter of [$\alpha$/Fe] is very important to determine the stellar age, in particular for the old stellar population. Moreover, no enough young stars meet our selection criteria in the catalog of \citet{Sanders2017}, e.g. only 299 stars are matched in the bin of $\tau<0.5$\,Gyr (the most left histogram). Therefore, we did not directly use the ages in this catalog, instead to re-calculate the ages with our procedure. Xiang et al. (in preparing) will discuss more about the stellar ages. Here, we just note that the age used in this study are basically consistent with ages of \citet{Sanders2017}, but exist systematical differences.

%That is why we only choose A-type stars with high temperature to obtain the plots in the young stellar bin.

\section{Conclusion} \label{sec:conclusion}

%% The reference list follows the main body and any appendices.
%% Use LaTeX's thebibliography environment to mark up your reference list.
%% Note \begin{thebibliography} is followed by an empty set of
%% curly braces.  If you forget this, LaTeX will generate the error
%% "Perhaps a missing \item?".
%%
%% thebibliography produces citations in the text using \bibitem-\cite
%% cross-referencing. Each reference is preceded by a
%% \bibitem command that defines in curly braces the KEY that corresponds
%% to the KEY in the \cite commands (see the first section above).
%% Make sure that you provide a unique KEY for every \bibitem or else the
%% paper will not LaTeX. The square brackets should contain
%% the citation text that LaTeX will insert in
%% place of the \cite commands.

%% We have used macros to produce journal name abbreviations.
%% \aastex provides a number of these for the more frequently-cited journals.
%% See the Author Guide for a list of them.

%% Note that the style of the \bibitem labels (in []) is slightly
%% different from previous examples.  The natbib system solves a host
%% of citation expression problems, but it is necessary to clearly
%% delimit the year from the author name used in the citation.
%% See the natbib documentation for more details and options.

In this Letter we reconstruct the spiral structures of \vR\ and \vPhi\ in the phase space of $z$-$v_z$ using LAMOST-Gaia combined stars. We further investigate the variations of the spiral pattern with different stellar ages, and find that (1) the spirals are gradually pronounced from $\tau<0.5$\,Gyr for both \vR\ and \vPhi, and then slowly become obscure until they almost disappear when $\tau>6.0$\,Gyr; (2)  Both \vR\ and \vPhi\ in the phase space have same wiggles in the different stellar ages bins, which suggest that the spirals probably have the same origin; (3) Different stellar populations respond to the perturbation by different levels of sensitivity. The stars in $\tau>6.0$ have no spirals for both \vR\ and \vPhi\ in the phase space, probably because the old stars are kinematically hot, and almost do not respond to the perturbation. We diagnose the stellar ages used in this study, and found our ages are systematically smaller by a fraction of 8\%$\sim$30\% than the ages of \citet{Sanders2017}.

According to the features of the observed spirals, we infer that the vertical perturbation to the disk took place no later than 0.5\,Gyr ago under the assumptions of (1) the spirals are sparked by an external perturbation, and (2) the perturbation on the ISM can be quickly dissipated, the newly born stars from the perturbed ISM do not share the general oscillations. The assumptions are possibly ideal. We need more observations or hydrodynamical simulations in high resolution to constrain the perturbation mechanism in the future.

 \acknowledgements
The authors thank Prof. James Binney for his very constructive comments, and thank Chervin F. P. Laporte, Robert Grand, Hao Tian and Ling Zhu for the helpful discussions. H.-J.T. acknowledges the National Natural Science Foundation of China (NSFC) under grants 11873034, 11503012, U1731124. C. L. acknowledges NSFC (Grant Nos. 11333003).
 Y. W. acknowledges NSFC (Grant No. 11403056). M.-S. Xiang acknowledges NSFC (Grant No. 11703035). This project was developed in part at the 2018 Gaia-LAMOST Sprint workshop, supported by the NSFC under grants 11333003 and 11390372. The Guo Shou Jing Telescope (the Large Sky Area Multi-Object Fiber Spectroscopic Telescope, LAMOST) is a National Major Scientific Project built by the Chinese Academy of Sciences. Funding for the project has been provided by the National Development and Reform Commission. LAMOST is operated and managed by National Astronomical Observatories, Chinese Academy of Sciences. This work has also made use of data from the European Space Agency (ESA) mission
{\it Gaia} (\url{https://www.cosmos.esa.int/gaia}), processed by the {\it Gaia}
Data Processing and Analysis Consortium (DPAC, \url{https://www.cosmos.esa.int/web/gaia/dpac/consortium}). Funding for the DPAC
has been provided by national institutions, in particular the institutions
participating in the {\it Gaia} Multilateral Agreement.

%% This command is needed to show the entire author+affilation list when
%% the collaboration and author truncation commands are used.  It has to
%% go at the end of the manuscript.
%\allauthors

%% Include this line if you are using the \added, \replaced, \deleted
%% commands to see a summary list of all changes at the end of the article.
%\listofchanges

\end{document}